\documentclass[twocolumn,showpacs,10pt,prb,floatfix,amssymb]{revtex4-1}

\usepackage[normalem]{ulem}
\usepackage{graphicx}
\usepackage{color}
\begin{document}

\newcommand{\nx}{\textrm}

\title{Quantum transport and observation of D’yakonov-Perel spin-orbit scattering in monolayer MoS$_2$}
	
\author{H. Schmidt$^{\dagger,1,2,*}$}
\author{I. Yudhistira$^{\dagger1,2}$}
\author{L. Chu$^{1,2}$}
\author{A. H. Castro Neto$^{1,2}$}
\author{B. \"Ozyilmaz$^{1,2}$}
\author{S. Adam$^{1,2,3,*}$}
\author{G. Eda$^{1,2,4,*}$}

\affiliation{$^1$Centre for Advanced 2D Materials and Graphene Research Centre, National University of Singapore, 6 Science Drive 2, Singapore 117546}
\affiliation{$^2$Department of Physics, National University of Singapore, 2 Science Drive 3, Singapore 117551}
\affiliation{$^3$Yale-NUS College, 16 College Ave West, Singapore 138527}
\affiliation{$^4$Department of Chemistry, National University of Singapore, 3 Science Drive 3, Singapore 117543}

\date{\today}
\begin{abstract}
Monolayers of group 6 transition metal dichalcogenides are promising candidates for future spin-, valley-, and charge-based applications. Quantum transport in these materials reflects a complex interplay between real spin and pseudo-spin (valley) relaxation processes, which leads to either positive or negative quantum correction to the classical conductivity. Here we report experimental observation of a crossover from weak localization to weak anti-localization in highly n-doped monolayer MoS$_2$. We show that the crossover can be explained by a single parameter associated with electron spin lifetime of the system. We find that the spin lifetime is inversely proportional to momentum relaxation time, indicating that spin relaxation occurs via D’yakonov-Perel mechanism.
 \end{abstract}

\maketitle
%
%
Quasi-two-dimensional (2D) crystals of group 6 transition metal dichalcogenides (TMDs) \cite{basic1,basic2,basic3} such as MoS$_2$ and WSe$_2$ have been recognized as a new class of semiconductors for spintronics and valleytronics\cite{spin, valley}. Due to distinct crystal symmetry and strong spin-orbit coupling, monolayer MoS$_2$ and other group 6 TMDs exhibit spin-split degenerate valleys at the corners (K and K' points) of the Brillouin zone. Since the spin and the valley degrees of freedom are coupled via time reversal symmetry, the valley degree of freedom can be accessed optically by circularly polarized light\cite{sflt,qwe}. A recent study has also shown that valley polarization can be electrically detected as anomalous Hall voltage arising from valley Hall effect\cite{valleyhall}. The exploitation of coupled spin and valley degrees of freedom is an intriguing approach to enabling novel spintronic and valleytronic device concepts\cite{valley}.\\

The use of spin- and valley-polarized charges as information carriers requires that the polarization state be preserved over a sufficiently long period. While recent experimental studies found the valley lifetime of optically generated excitons to be on the order of nanoseconds\cite{sflt}, little is known about the relaxtion lifetime in unipolar charge transport. In this regard, quantum transport\cite{theory_1,theory3} has been suggested as an effective probe to study the dynamics of scattering processes that lead to loss of spin and valley polarization.\\

Unlike the Drude-Boltzmann semi-classical transport, the quantum corrections to the conductivity are interference effects, and are therefore universal — in the sense that they should not depend on the details of the microscopic mechanisms at play. However, as discussed in literature \cite{intro1}, there is a long tradition of extracting information about the underlying microscopic mechanisms from the quantum transport. For example, in GaAs heterostructures the quantum interference correction to the classical conductivity is determined by the breaking of spin-rotational symmetry by spin-orbit coupling \cite{hln}. But since the spin-relaxation rate changes with carrier density, Miller et al. \cite{intro2} observed a crossover from pure weak localization (WL) at low carrier density to pure weak anti-localization (WAL) at high carrier density. A similar phenomenon has been explored in graphene. Unlike GaAs, for graphene, it is not the spin degree of freedom that is important, but the intervalley scattering that can be represented as a breaking of pseudospin-rotational symmetry \cite{intro3}. By exploring this crossover caused by breaking inversion symmetry, Refs. \onlinecite {intro4,intro5} showed, for example, that the intervalley scattering in graphene comes from the edges of the graphene ribbons and not from the bulk. As these examples illustrate, quantum transport can nonetheless provide important information about the microscopic mechanisms at play.\\

\begin{figure*}[t]
\includegraphics[width=1.95\columnwidth]{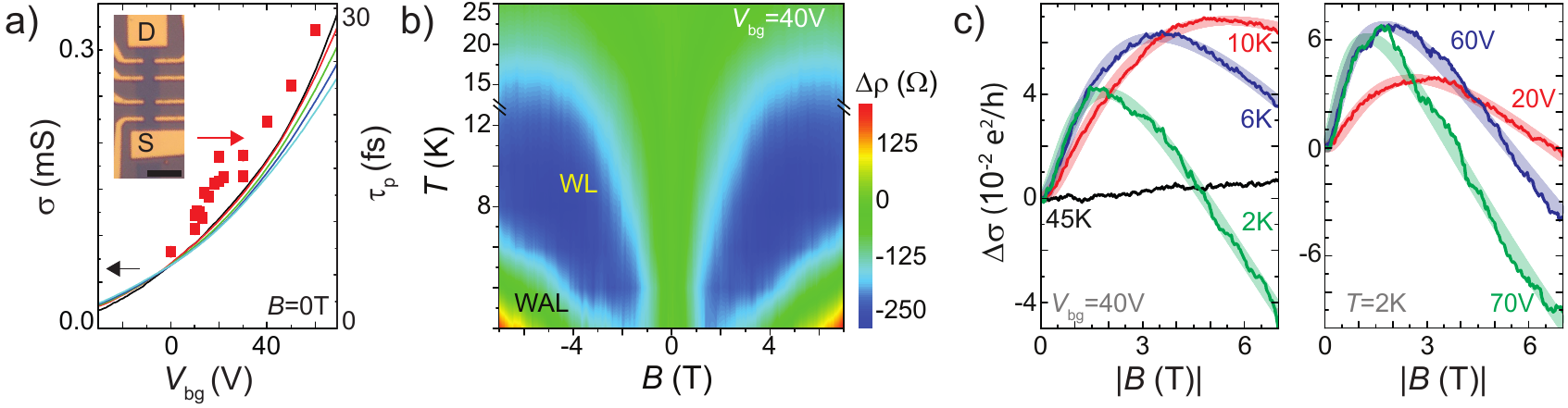}
\caption{\label{fig:bild1} a) Transfer curves measured at different temperature: 20, 40, 60, 80, 100 K from black to cyan. The red squares represent $\tau_p$ at $T=2~\mathrm{K}$. Inset: optical microscope image of a rectangular device with current source (S) and drain (D) contacts (scale bar: 5~$\mu$m). b) Experimentally observed MR as a function of magnetic field and temperature measured at $\mathrm{V}_{bg} = 40~\mathrm{V}$ and $\mathrm{V}_{tg} = 1\mathrm{V}$. c) $\Delta\sigma$ at $T=$ 2, 6, 10, 45~K and a fixed backgate voltages of $V_\mathrm{bg}$=40~V (left panel) and at fixed temperature of $T$ = 2~K and different gate voltages of $V_\mathrm{bg}$= 20, 60, 70~V (right panel). The solid lines indicate experimental data, and the shaded lines the according fits using Equation (2). To calculate $\Delta\sigma$, the initial measured MR curve was symmetrized to avoid contributions from the sample geometry, so that $\rho(B)=(\rho(+B)+\rho(-B))/2$.}
\end{figure*}
Although different mechanisms for spin and valley dynamics in MoS$_2$ have been studied theoretically \cite{theory3,IV-scat,theory4,theory5,theory6}, purely electronic experiments on monolayers have thus far been missing. Previous quantum transport studies have been limited to measurements on multilayers\cite{zeeman, scdome,phase}, in which coupled spin and valley physics is absent. On the other hand, experiments on monolayers have mainly focused on basic charge transport\cite{basic2, intrinsic, wujing}. In this Letter, we report experimental observation of the crossover from WL to WAL in highly n-doped monolayer MoS$_2$. We show that, in the limit of large separation of length scales, Hikami-Larkin-Nagaoka (HLN) approach can be used to extract the spin-relaxation time $\tau_\mathrm{SO}$ from the magneto-conductivity (MC). Our analysis reveals that $\tau_\mathrm{SO}$ is inversely proportional to the momentum relaxation time $\tau_p$ as one would expect for D’yakonov-Perel (DP) mechanism. This dominance of DP spin relaxation is consistent with recent theoretical expectations \cite{theory4,theory5}. Moreover, we find that the dominant form of phase-decoherence is electron-electron (e-e) scattering which is expected in monolayer MoS$_2$   where the interaction strength is at least a factor of 10 larger compared to conventional 2D electron gases\cite{intro1}.\\

%
%
Our experiments were conducted on dual-gated mechanically exfoliated monolayer MoS$_2$ on a SiO$_2$/p-Si substrate as previously reported \cite{chu}. This gating technique is helpful in studying charge transport in the high carrier density regime where conduction occurs via extended states. All magneto-transport measurements were made for conditions in the diffusive transport regime i.e. $\sigma \gtrsim e^2/h\approx 0.04~\mathrm{mS}$ (Fig. 1a). From the Hall signal, carrier densities on the order of $n_\mathrm{Hall}\approx 10^{13}~\mathrm{cm}^{-2}$ and Hall-mobilities of $\mu_\mathrm{Hall}\approx 130~$cm$^2$V$^{-1}\mathrm{s}^{-1}$ at low temperatures were obtained, similar to earlier reports \cite{intrinsic,mobeng}. It is worth noting that clean Hall signals were obtained only at sufficiently high carrier densities where metallic conduction was observed, possibly due to vanishing effects from localized states \cite{CVD2} and Schottky barriers. Mean free path $\ell$ and momentum relaxation time $\tau_\mathrm{p}$ were obtained using $n_\mathrm{Hall}$ and $\mu_\mathrm{Hall}$. In our experimental window, $\tau_\mathrm{p}$ was found to increase gradually with gate voltage. Thus, we use the gate bias as the knob to continuously tune $\tau_\mathrm{p}$ (Fig. 1a).\\

An overview of our experimental results is shown in Fig. 1b which depicts the as-measured magneto-resistivity (MR), $\Delta \rho (B) = \rho(B)-\rho_0$, where $\rho_0$ is the zero field resistivity, as a function of magnetic field and temperature at a charge carrier density of $n\approx1.5\times10^{13}~\mathrm{cm}^{-2}$. While no significant MR is observed above $T=20$~K, the measurements at lower temperatures show two distinct regimes. At small magnetic fields, the MR is negative, indicating dominance of WL, i.e. negative correction to the classical conductivity (blue region in Fig. 1b). In contrast, at $T <$ 6~K and $B>$ 3~T, the MR changes sign, revealing prevailing WAL, i.e. positive correction to the classical conductivity (Fig. 1c left). The behavior of WL-WAL crossover was also affected by gate voltages (Fig. 1c right). With increasing back gate voltage, the crossover point shifted towards lower magnetic fields. We found that the general trends were similar in bilayer samples\cite{chu_wl}. Note that this WL-WAL crossover behavior is just the opposite to what has been observed in conventional 2D electron gases\cite {intro2,Thillosen06}. Our results also differ from earlier reports on bulk MoS$_2$ where no crossover was observed with temperature \cite{phase,zeeman,scdome}. This observation of WAL at fields of $B \sim 3~\mathrm{T}$ and above is somewhat surprising in light of recent theoretical papers \cite{IV-scat,theory_1} which predict the exact opposite, i.e. WAL at low fields and WL at larger fields. The disappearance of positive (negative) MR for $T/T_{\rm F}\approx 0.02$~(0.1), where $T_\mathrm{F}$ is the Fermi temperature, rules out semi-classical effects as the origin of the observed MR at low temperatures. Further, the MR curves do not collapse according to Kohlers rule (MR $\propto f(\mu$B)), indicating that our observations cannot be explained by classic quadratic background $\propto \mu^2B^2$ that was observed in multilayer WSe$_2$\cite{zeeman}.\\

%
%
We propose two possible explanations for the discrepancy between our observations and theoretical predictions. First, it is possible that the conduction bands have some spin-texture giving rise to a $\pi$-Berry phase. Indeed, this model for the conduction bands was recently invoked to explain the observation of valley Hall effect in monolayer MoS$_2$ \cite{valleyhall}. However, for the range of carrier densities used in our experiment, we have $E_\mathrm{F} << \Delta$ with the bandgap $\Delta\approx 2~\mathrm{eV}$, resulting in a negligible Berry's phase and a conserved sublattice isospin\cite{theory_1,theory3}. The second possible explanation is separation of length scales. The system is characterized by spin, valley, and sublattice degrees of freedom contributing to the quantum transport \cite{theory_1}, and it is conceivable that in a realistic experiment, there is a large separation between different length scales and scattering rates. For example, any scattering length scale that is larger than phase-coherence length cannot be probed within our experiment. Similarly, if the effective magnetic field corresponding to a particular microscopic scattering mechanism is much larger than the largest magnetic fields $B_\mathrm{max}$ that we measure, our experiment would not be able to probe its presence.
\begin{figure}[t]
\includegraphics[width=.9\columnwidth]{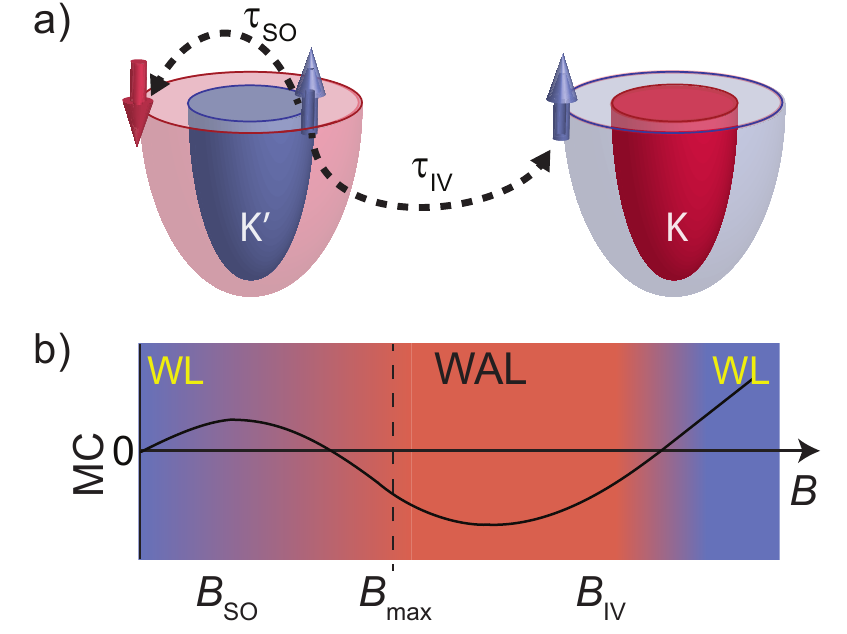}
\caption{\label{fig:bild3} a) Intra- and Intervalley scattering in the spin-split K and K' valleys of the conduction band. b) A sketch of predicted MC behavior assuming large separations in the time scales of relevant scattering mechanisms. Two parameters $B_\mathrm{SO}$ and $B_\mathrm{IV}$ (and accordingly $\tau_\mathrm{SO}$ and $\tau_\mathrm{IV}$) define the crossover between WL and WAL. }
\end{figure}\\

Formally, generic disorder that couples spin, valley and sublattice degrees of freedom in monolayer MoS$_2$ is described by $U(8)$ algebra with $8^2 = 64$ generators. This implies that there are $64$ different channels with corresponding relaxation rates that couple the eigenstates of the clean system. The problem can be significantly simplified by symmetry arguments and in the limits of large separation between length scales. Our experiments show a single crossover from WL to WAL, which implies that we should be able to construct a theory with only one Cooperon relaxation mode. Moreover, the fact that we observe WL at low magnetic fields implies that the triplet channel from either the spin $SU(2)$ mode or the valley $SU(2)$ mode is gapped in our experiment.\\

To proceed, we estimate some of the relevant scattering rates based on our measurements and previous reports. It has been reported that MoS$_2$ contains significant density of sulfur vacancy defects. Recent electron microscopy measurements estimated the density to be $ 10^{13} \mathrm{cm}^{-2}$~~\cite{vacancies,hong}. This defect density is also consistent with our analysis of DC transport measurements\cite{me} and looking at the scaling of mobility with carrier density, dielectric constant, and temperature for calculations using different combinations of impurities \cite{jena2014}. The different estimates of the defect density are consistent to within $\pm$ 20 percent and translate to a spin-conserved intervalley scattering length of $L_{\rm IV} \approx 3~{\rm nm}$ and corresponding magnetic field of $B_{\rm IV}=\hbar/4eL_\mathrm{IV}^2 \gtrsim 16~{\rm T}$. This is clearly larger than the maximum magnetic field of our experiment $B_{\rm max}$. On the other hand, spin-orbit mediated spin-relaxation in monolayer MoS$_2$ has been predicted to occur at time scales of several pico-seconds in the conduction band$^{21}$, corresponding to a spin-orbit scattering length of $L_{\rm SO} \approx 10~{\rm nm}$, or a magnetic field of $ B_{\rm SO} \approx 2~{\rm T}$ which is within our experimental window. Thus, our experimental system is subject to the following separation of scales,
\begin{equation}
L_{\phi} \gtrsim L_{\rm SO} > L_{\rm Bmax} \gtrsim L_{\rm IV} > \ell 
\end{equation}
where $\ell$ is the mean free path, which is on the same order of magnitude as $L_{\rm IV}$ based on our transport measurements. It is now clear that for the range of experimental magnetic fields, time-reversal symmetry is not completely broken, but the large intervalley scattering breaks our pseudospin rotational symmetry, giving us the symplectic universality class for the valley $SU(2)$, and thus WAL. The observed crossover is therefore characterized by another symmetry breaking process, spin flip scattering, which leads to a change in the universality class as sketched in Fig.~2b.\\

In the corresponding limit that $B_{\rm IV}\gtrsim B_{\rm max} >> B_{\rm SO}$, we simplify the problem by assuming that all the valley triplets are gapped, and keep only the 4 Cooperons for the spin $SU(2)$ degrees of freedom. This result follows in a straight-forward manner from the seminal paper by Hikami, Larkin and Nagaoka \cite{hln}. We find
\begin{widetext}
\begin{eqnarray}
\Delta \sigma &=& \frac{e^2}{2\pi h} \left [  F \left( \frac{B}{B_\phi}\right)  - 2 F \left( \frac{B}{B_\phi + B_{\rm SO}^x + B_{\rm SO}^z}\right) -  F \left( \frac{B}{B_\phi + 2 B_{\rm SO}^x}\right) \right], \nonumber \\
&=& \frac{e^2}{2\pi h} \left [  F \left( \frac{B}{B_\phi}\right)  - 3 F \left( \frac{B}{B_\phi + 2B_{\rm SO}} \right) \right]. \label{Eq:main}
\end{eqnarray}   
\end{widetext}
\noindent where $F(z)$ is defined by $F(z) = \ln(z) + \psi\left(\frac{1}{2}+\frac{1}{z}\right)$ with the digamma function $\psi$. In the last line we assumed that each of the spin-triplet Cooperons have the same relaxation time. Another choice would have been to set $B_{\rm SO}^x = B_{\rm so}^y =  B_{\rm SO}$ and $B_{\rm SO}^z = 0$, but within the resolution of the experiment, this would give identical results for $\tau_{\rm SO}$. Quite generally, Eq.~\ref{Eq:main} represents a generic crossover from the orthogonal to the symplectic universality class parametrized by only two terms $B_{\rm \phi}$ and $B_{\rm SO}$. Here, $B_\phi^{-1}= 4 e L_{\rm \phi}^2/\hbar$ is inversely related to the phase-coherence length $L_{\rm \phi}$ and $B_{\rm SO}^{-1} = 4e D \tau_{\rm SO}/\hbar$ measures a spin-orbit scattering mechanism with relaxation time $\tau_{\rm SO}$, where $D$ is the classical Drude diffusion constant. As shown in Fig.~1c, our severely constrained two-parameter fit yields excellent agreement with our experimental data, implying that only spin-orbit scattering is relevant in our experimental window.
\begin{figure}[t]
\includegraphics[width=.9\columnwidth]{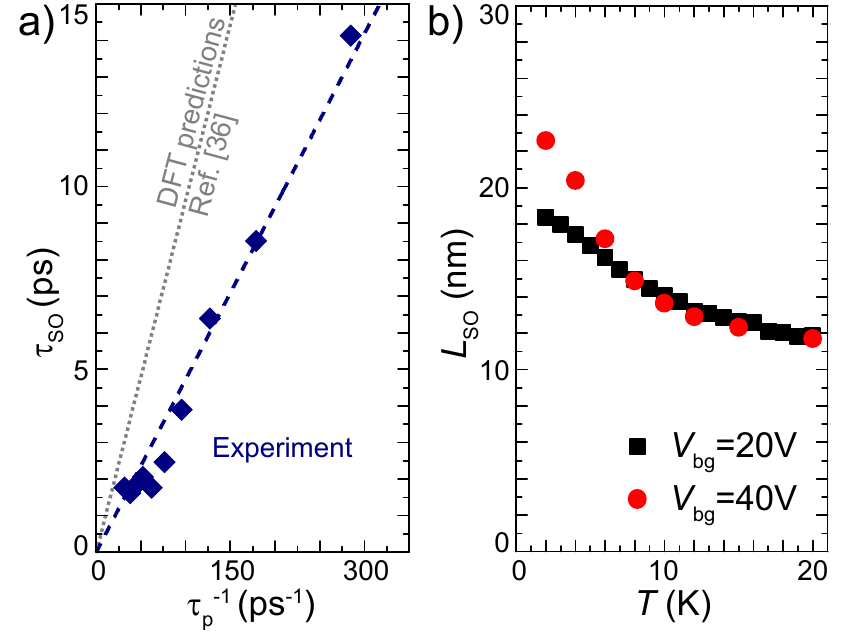}
\caption{\label{fig:bild4} a) The crossover parameter $\tau_\mathrm{SO}$ as a function of $\tau_\mathrm{p}^{-1}$. The blue dashed line is a linear fit and the grey dashed line is a theoretical prediction based on $\lambda_{int}$=3meV [\onlinecite{DFT}]. b) $L_\mathrm{SO}$ as a function of temperature.}
\end{figure}\\

%
%
Fig. 3 shows $\tau_\mathrm{SO}$ and $L_\mathrm{SO}$ as a function of inverse momentum relaxation time $\tau_\mathrm{p}^{-1}$ and temperature, respectively. Note that the spin lifetime we discuss here is different from those obtained by optical pump probe experiments, where Coulomb interaction between electrons and holes play a dominant role in the scattering processes\cite{lifetimes}. We find that $\tau_\mathrm{SO} \propto \tau_\mathrm{p}^{-1}$ and $\tau_\mathrm{SO} >> \tau_{p}$, which is what one would expect for the DP spin relaxation mechanism. We can also use the relationship $2 \hbar^2/(\tau_p \tau_\mathrm{SO}) = \lambda_\mathrm{int}^2$ to estimate the strength of the spin-orbit interaction $\lambda_\mathrm{int}$. From Fig.~3a we find $ \lambda_\mathrm{int} \approx 4.3\pm 0.1$~meV, which is within the range expected from DFT calculations \cite{DFT}. We also find that the magnitude of $L_\mathrm{SO}$ is only weakly dependent on density and temperature. Since $L_\mathrm{SO}=\sqrt{(1/2)v_F^2 \tau_\mathrm{p} \tau_\mathrm{SO}}$, the results are consistent with the expectation that the spin-orbit interaction strength is independent of temperature. The weak temperature and density dependence of $L_\mathrm{SO}$ is also an indication that other spin relaxation mechanisms such as Elliot-Yaffet and Bir-Aronov-Pikus scattering are not dominant in our system, and provide additional evidence for DP spin relaxation.
\begin{figure}[t]
\includegraphics[width=.9\columnwidth]{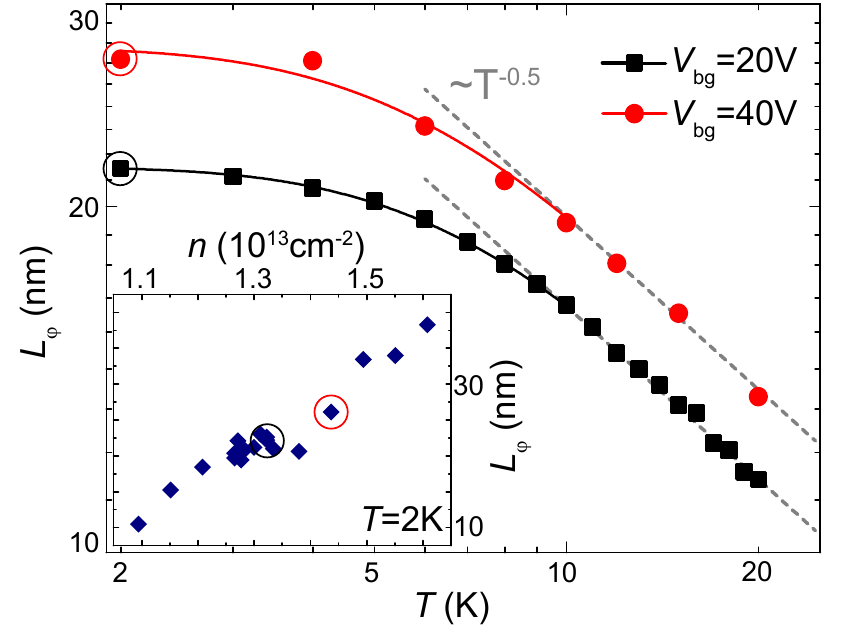}
\caption{\label{fig:bild5} Temperature dependence of $L_{\phi}$ at $V_\mathrm{bg}=$20 V and 40 V with guide to the eyes (solid lines). The dashed lines show agreement with $\propto T^{-0.5}$ dependence for $T>$10~K. Inset: density dependence of $L_{\phi}$ at $T=2$~K.}
\end{figure}\\

%
%
Finally, we analyze the phase coherence of our system. Figure 4 depicts the phase coherence length $L_\mathrm{\phi}=\sqrt{\hbar/4eB_\phi}$ for two different gate voltages as a function of temperature. For both cases, the data shows $L_\phi\propto T^{-\alpha}$ behavior for higher temperatures with $\alpha$ on the order of 0.5, indicating that e-e interaction limits the phase coherence in this regime. Below 10~K, a saturation is observed similar to other 2D systems \cite{graphenesat}, pointing towards an additional dephasing mechanism. Nevertheless, we find that the phase coherence length shows linearly increasing trend with charge carrier density (inset Fig.~4), which is consistent with theoretical prediction for dephasing due to e-e scattering\cite{ee}, even in the saturation regime.\\

%
%
In summary, we have studied low temperature quantum electron transport in monolayer MoS$_2$. The crossover between WL and WAL in this system indicates a separation of relevant length scales due to the high concentration of short range scatterers. We describe this crossing with a single parameter, which is associated with intravalley spin-flip scattering. The scattering time clearly shows the signatures of DP relaxation as predicted by theory. Further, the phase coherence length is found to be limited by e-e interaction at temperatures above 10 K.\\

The authors would like to thank Hector Ochoa for fruitful discussions. This research is supported by the National Research Foundation, Prime Minister’s Office, Singapore under its Medium sized centre programm as well as the grant NRF-NRFF2011-02 (G.E.) and NRF-NRFF2012-01 (S.A.), by the Singapore Ministry of Education and Yale-NUS College (R-607-265-013121 (S.A.)), the NRF-CRP award 'Towards Commercialization of Graphene Technologies' (R-144-000-315-281 (B.O.)) and the NRF-CRP award 'Novel 2D materials with tailored properties: beyond graphene' (R-144-000-295-281 (A.H.C.N.)).

\newpage
\end{document}